\documentclass[aps,prd,preprint,groupedaddress]{revtex4}
\usepackage{latexsym,amsfonts}
\usepackage{hyperref}
\usepackage{graphicx}
\usepackage{epstopdf}
\begin{document}     

\title{The Field Theory of Specific Heat\footnote{\href{http://hdl.handle.net/11858/00-001M-0000-002A-2B37-9}{Russian Journal of Mathematical Physics, {\bf 23} (1), 56-76 (2016)}}
\footnote{followed by \href{https://arxiv.org/abs/1801.03241}{arxiv:1801.03241}}}
\author{Yuri V. Gusev}

\affiliation{Lebedev Research Center in Physics, Leninskii Prospekt 53, Moscow 119991, Russia,  {\tt yuri.v.gussev@gmail.com}}
\affiliation{Max Planck Institute for Gravitational Physics (Albert Einstein Institute), Am M\"uhlenberg 1, D-14476 Golm, Germany}
\date{April 6, 2015}

\begin{abstract}
Finite temperature quantum field theory in the heat kernel method is used to study the heat capacity of condensed matter. The lattice heat is treated a la P. Debye as energy of the elastic (sound) waves. The dimensionless functional of free energy is re-derived with a cut-off parameter and used to obtain the specific heat of crystal lattices. The new dimensionless thermodynamical variable is formed as the Planck's inverse temperature divided by the lattice constant. The dimensionless constant, universal for a class of crystal lattices, which determines the low temperature region of molar specific heat, is introduced and tested with the data for diamond lattice crystals. The low temperature asymptotics of specific heat is found to be the fourth power in temperature instead of the cubic power law of the Debye theory. Experimental data for the carbon group elements (silicon, germanium) and other materials decisively confirm the quartic law. The true low temperature regime of specific heat is defined by the surface heat, therefore, it depends on geometrical characteristics of a body, while the absolute zero temperature limit is geometrically forbidden. The limit on a growth of specific heat at temperatures close to critical points, known as the Dulong-Petit law, appears from the lattice constant cut-off. Its value depends on the lattice type and it is the same for materials with the same crystal lattice. The Dulong-Petit values of compounds are equal to those of elements with the same crystal lattice type, if one mole of solid state matter were taken as the Avogadro number of the lattice atoms. This means the Neumann-Kopp rule is valid only in some cases.
\end{abstract}
\maketitle

\section{Quantum field theory for condensed matter}      \label{QFT}                                        

When developing a physical theory of condensed matter, e.g., solid state
matter, one usually starts with the atomistic view of matter, which consists of
particles, e.g., atoms (ions) of the lattice and electrons
\cite{Kittel-book1996,Landau-V5}. However, particles and quasiparticles in a
physical theory are employed only at intermediate stages, while final
expressions are some static functions, which describe integral properties of
condensed matter systems, such as the heat capacity, electric conductivity, etc.
One usually does not intend to and cannot describe the behaviour of individual
particles.  Furthermore, one normally starts with a system of non-interacting
constituents, e.g., independent oscillators of A. Einstein
\cite{Einstein-AdP1907,Kittel-book1996} or free electron gas of P. Drude
\cite{Drude-AdP1900,Kittel-book1996}, and later builds in interactions between
them to develop eventually physics of 'strongly interacting' (correlated)
systems. Building a theory this way follows the historical path of physics
development and therefore relies on thermodynamics and statistical mechanics of
{\em gases}. However, physical properties of solid and liquid matter are very
different from those of gases. Indeed, a condensed matter system can be defined
as a system of particles that form physical {\em continuum}. In fact, the view
on condensed matter as continuous medium is well developed in such classic
branches of physics as elasticity theory \cite{Landau-elastV7,Love-book1944,Royer-book2000} and hydrodynamics \cite{Landau-hydroV6}. These are
phenomenological theories, which are not concerned with particle compositions of
condensed matter systems, but they are capable of describing matter's properties
by summarizing them as physical laws.  Elasticity theory describes mechanical
phenomena of solid state matter, yet its mathematical apparatus can be used for
the theory of thermal phenomena, as was first proposed  by P. Debye
\cite{Debye-AdP1912}.

Temperature is a key variable of any thermal theory. Statistical thermodynamics
\cite{Landau-V5} was developed as physics of gas, which is viewed as a
collection of massive particles moving in free space. Correspondingly, this
physical theory used classical mechanics as its foundation. In statistical
physics temperature is an emergent characteristic of a large number of
particles. It appeared possible to use {\em mechanics} of particles (system's
constituents) to find a {\em statistical} distribution of their kinetic energies
(velocities), which corresponds to an observed temperature. This was done first
by J.C. Maxwell \cite{Maxwell-book2011}. Since the physical nature of
temperature in a condensed matter system, which presents medium in a bounded
space, is different, this procedure cannot work. In a quantum theory of matter
temperature is introduced as an external (axiomatic) parameter. This is also the
approach of the present work, but it is implemented in a geometrical language, since we
believe that thermal and electronic properties of condensed matter can be better
described by the methods of field theory and geometry.

In statistical mechanics, in order to derive macroscopic thermal quantities, the
number of particles is taken to infinity, while the volume occupied by particles
is taken to infinity as well \cite{Landau-V5}. Both limits are unphysical, but
sought physical quantities are derived finite. In this limit, the
influence of a system's boundaries on its physical properties is infinitesimal
and cannot be studied.  However, finite size effects, e.g., in heat conductivity
and heat capacity, are observed, usually at low absolute temperatures. For
some condensed matter systems under certain conditions, effects of the body's
geometrical characteristics can become the leading. The Schwinger-DeWitt (geometric) formalism for quantum field theory (QFT)  \cite{DeWitt-book2003,GAV-lectures2007}
that we use naturally incorporates boundary effects due to its mathematical
foundation, the heat kernel. 

We will use the finite temperature QFT \cite{Gusev-FTQFT2015} to develop a theory of specific heats of crystal lattices that may also be applicable to other forms of condensed matter. Let us first recall its basic ideas. Since any condensed matter system exists within a certain three-dimensional domain, and its behaviour in time is not usually studied, the field theory can be defined in the Euclidean four-dimensional spacetime, i.e., the spacetime metric's signature is not Minkowskian, but Euclidean (no time coordinate is singled out). The physical time is imaginary, and it is distinguished from other spacetime dimensions only by its closed topology, $\mathbb{S}^1$. The orbit's length of the closed Euclidean time is expressed via the fundamental physical constants as the Planck's inverse temperature, 
\begin{equation}
\beta = \frac{\hbar v}{k_B T}, \label{beta}
\end{equation}
where $k_B$ is the Boltzmann's constant and $\hbar$ is the reduced Planck's
constant. Note, that in contrast to the definition in \cite{Gusev-FTQFT2015},
there is no calibration coefficient in (\ref{beta}) because temperature $T$, or $\beta$ for
that matter, is not an independent variable in the present setting.  The
characteristic velocity, $v$, enters the definition (\ref{beta}). For the
electronic component of heat it is supposed to be the speed of light, while for
the elastic (acoustic) component, contributed by the crystal lattice, it is the
velocity of elastic (sound) waves. 	

In the Schwinger-DeWitt quantum field theory, the Laplace operator,
\begin{equation}
\Box= g^{\mu\nu} \nabla_{\mu} \nabla_{\nu},	\label{operator}  
\end{equation}
defines a particular field model \cite{BarVilk-PRep1985}. It is constructed of
the covariant derivatives, which may contain the metric (gravity) and gauge
field (e.g., electromagnetic) connections. For the theory of lattice heat, gauge fields and
gravity are not relevant, and the Laplacian (\ref{operator}) is trivially the
second order partial derivatives. For the study of electronic properties,
electromagnetic fields cannot be neglected, in particular, they should be
important for the electronic specific heat. 

The kernel of the heat equation \cite{DeWitt-book2003,GAV-lectures2007,BarVilk-PRep1985},
\begin{equation}
	\Big(\frac{\partial}{\partial s} - \Box^x \Big) K (s| x,x') = 0,\ 
	K (s| x, x')|_{s\rightarrow 0} = \delta(x,x'),         \label{heateq}
\end{equation} 
is expressed via the proper time parameter, $s$. The trace of the kernel in a compact three-dimensional manifold with boundary  is \cite{Gusev-FTQFT2015},
\begin{equation} 
	{\mathrm{Tr}} K(s) 
	\equiv 
	\int {\mathrm d}^{3} x \,   
	K(s|x,x) 
	 = 
	\frac1{(4\pi s)^{3/2}}\ 
	\mathcal{V}
 	+ \frac1{(4\pi s)}\
 	\mathcal{S},  \label{TrKlocal} 
\end{equation}
where $\mathcal{V}$ is the volume and $\mathcal{S}$ is the boundary's area of
the bounded domain of  manifold $\mathbb{R}^3$. This expression is valid at an
arbitrary proper time. For the reason explained in Sect.~\ref{size} we omit the
boundary ('surface') term and keep only the volume ('bulk') term of this
expression.

Let us compute the functional of free energy, defined in the usual way
\cite{Gusev-FTQFT2015}, but with the positive lower limit of the the proper time
integral. This is the standard QFT regularization used to remove the ultraviolet
divergences \cite{BarVilk-PRep1985}. Our integral is finite, but the lower
limit appears due to the wavelength cut-off caused by discreteness of the
lattice. Because the length square, $\tilde{a}^2$, is proportional to the proper time,
$s$, which parametrizes the geodesic \cite{DeWitt-book2003,BarVilk-PRep1985}, we
introduce the integral's lower limit as (the numerical coefficient is choosen to
simplify expressions below),
\begin{equation}
-F^{\beta}_{\tilde{a}} \equiv {\tilde{A}}
 \int_{\tilde{a}^2/4}^{\infty}\! \frac{{\mathrm d} s}{s}\,  
	{\textrm{Tr}}  K^{\beta}(s).         \label{Fbeta}
\end{equation} 
An observable quantity, such as specific heat, must certainly be independent
of the regulator (cut-off parameter), and in Sect.~\ref{specificheat} we show
that indeed it is. The overall numerical coefficient ${\tilde{A}}$ is to be
calibrated by experiments. We substitute the three-dimensional heat kernel trace
(\ref{TrKlocal}) into the thermal (3+1)-dimensional expression
\cite{Gusev-FTQFT2015},
\begin{equation} 
	{\textrm{Tr}}{K}^{\beta}(s)=  
	\frac{\beta}{(4 \pi s)^{1/2}}\
	\sum_{n=1}^{\infty} {\mathrm{e}}^{-\frac{\beta^2 n^2}{4s}}  \
	{\mathrm{Tr}} K(s).  \label{TrKbeta}
\end{equation}
After applying the change of variables, $y =\beta^2/(4 s)$, free energy (\ref{Fbeta}) looks like
\begin{equation}
	-F^{\beta}_{\tilde{a}} = 
	\frac{{\tilde{A}}}{\pi^2}\, 
	\frac{\mathcal{V}}{\beta^3}
	\sum_{n=1}^{\infty}\, \int_{0}^{\alpha^2}{\mathrm{d}} y   \,
	  y \,
	   \mathrm{e}^{-y n^2}. \label{Fba}
\end{equation}
It is only different from the previously studied expression \cite{Gusev-FTQFT2015} by  the integral's upper limit, which is expressed as a new {\em dimensionless} variable,
\begin{equation}
\alpha \equiv \frac{\beta}{\tilde{a}} =  \frac{\hbar v}{\tilde{a} k_B  T}.
\label{alpha}
\end{equation}
The computed expression (\ref{Fba}) accepts the final form,
\begin{equation}	   
	-F^{\alpha} = 
	\frac{{\tilde{A}}}{\pi^2}\,
	\frac{\mathcal{V}}{\tilde{a}^3}\, 
	\sum_{n=1}^{\infty}\, \frac{1}{n^4 \alpha^3}\Big(1 -\exp (-\alpha^2 n^2) 
	- n^2 \alpha^2 \exp (-\alpha^2 n^2)\Big). \label{FDP}
\end{equation}
This sum's first term is the zeta function,  $\zeta(4) = {\pi^4}/{90}$.
The two other terms cannot be analytically computed.

The new upper index of the functional $F^{\alpha}$ indicates a
change in physical understanding of the free energy. We declare the
dimensionless variable $\alpha$  be {\em the thermal variable for condensed matter systems}, i.e., field theory  models with the short wavelength cut-off, that supersedes old variables, the absolute temperature, $T$ (Kelvin), and the Planck's inverse temperature, $\beta$ (meter). Indeed, it is natural that the dimensionless functional depends on the dimensionless variable; the theory could be called conformal (or scale free) in the popular language.

\begin{figure}[h!]
  \caption{Function $\Theta(\alpha)$ for the finite sum, n=400}
  \centering
\includegraphics[scale=0.5]{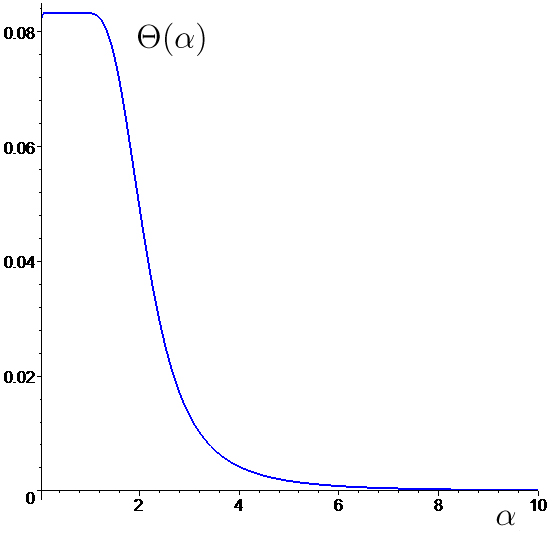}
\label{thetaalpha}
\end{figure}

Following the line of
\cite{Gusev-FTQFT2015}, we take the derivative of (\ref{FDP}) over the
 variable $\alpha$ to obtain,
\begin{equation}	   
	\frac{\partial F^{\alpha}}{\partial \alpha} = 
	\tilde{A} \frac{3}{\pi^2} \frac{\mathcal{V}}{\tilde{a}^3}\, 
	\Theta(\alpha), \label{derivative}
\end{equation}
where the notation for the dimensionless thermal sum is introduced,
\begin{equation}
\Theta(\alpha) =
\sum_{n=1}^{\infty}\,
	\frac{1}{ n^4 \alpha^4} 
	\Big\{
	 1 -\exp (-\alpha^2 n^2)
	 - n^2 \alpha^2\,  \exp (-\alpha^2 n^2)
	- \frac{2}{3} n^4 \alpha^4 \,  \exp (-\alpha^2 n^2)
	\Big\}. \label{Theta}
\end{equation}
We conjecture that  $\Theta(\alpha)$ is a {\em  universal} function of
temperature scaling for condensed matter systems.
The plot of $\Theta(\alpha)$, computed in Maple for the finite sum $n=400$, is
shown in Fig.~(\ref{thetaalpha}).
The  maximum value of this quantity is found to be,
\begin{equation} 
\Theta_{\mathrm{max}} \approx 8.33 \cdot 10^{-2}. \label{Thetamax}
\end{equation}
It might be convenient to redefine (\ref{Theta}) by $10^2$ especially because it usually has a rather large dimensionful factor, (\ref{Rsolid}), see Sect.~\ref{DPvalues}.
Three other terms of the sum have good behaviors in asymptotics, $\alpha
\rightarrow 0$ ('high temperature') and $\alpha \rightarrow \infty$ ('low
temperature'). The limit $\alpha=0$ is topologically forbidden  since it would
be equivalent to the open manifold, $\mathbb{R}^1$, \cite{Gusev-FTQFT2015}. Even
though a numerical estimate for the expression (\ref{derivative}) becomes
negative, when $\alpha$ goes to zero (note, the plot in Fig.~\ref{thetaalpha} is
from $\alpha=0.05$), its intermediate {\em asymptotics} is positive constant for
any infinitesimal value of $\alpha$, when the sum's upper limit tends to
infinity. In fact, the leftmost bound of $\Theta(\alpha)$ is irrelevant,
as in the regime, $\beta \ll a$ (the thermal wavelength of sound is much less
than the lattice cut-off), the theory breaks down. 

The expression (\ref{derivative}) is valid at any values of its
variables. We consider now expansions of this functional in its
dimensionless thermal variable. The 'low temperature' asympotics is easily
found, 
\begin{equation}	   
	\frac{\partial F^{\alpha}}{\partial \alpha} = 
	{\tilde{A}}
	\frac{\pi^2}{30}\,
	\frac{\mathcal{V}}{\tilde{a}^3}\, \frac{1}{\alpha^4}, \ \alpha \rightarrow 
\infty . \label{lowalpha}
\end{equation}
The 'high temperature' asymptotics can be explored numerically. The
maximum value of $\Theta(\alpha)$ gives us the constant term,
\begin{equation}	   
	\frac{\partial F^{\alpha}}{\partial \alpha} = 
	\tilde{A} \Theta_{\mathrm{max}} \frac{3}{\pi^2}
\frac{\mathcal{V}}{\tilde{a}^3}, \ \alpha  \rightarrow 0. \label{highalpha}
\end{equation}
These simple derivations are quite general and can hopefully be used for
different problems of thermal physics.

\section{The field theory of specific heat}      \label{theory}        
                               

\subsection{From elasticity theory to the heat kernel} \label{elasticity}  
                                     

Let us apply this formalism to the physics of lattice heat of solid state
matter.  The theory we  want to build can be viewed as a modified completion
of the theory of specific heat of Peter Josef William Debye
\cite{Debye-AdP1912}. One century after its creation, his theory is still widely used in practice. The key idea of P. Debye is that the lattice heat is energy of
the standing sound waves in a solid body. The sound waves are elastic waves
\cite{Love-book1944}, therefore, the velocity of sound in a solid body
(crystal lattice), which is typically of the order of $10^3\ m/s$,  enters the
Planck's inverse temperature (\ref{beta}) and correspondingly
the $\alpha$-variable  (\ref{alpha}).
This velocity's magnitude lets us leave aside the electronic contribution,
because it is separated from the lattice heat by the factor, $v/c \approx
10^{-5}$, taken to some power. 

Free energy of finite temperature QFT (\ref{FDP}) is applicable to this
problem because the operator of the equations of elastic (sound) waves
accepts the required form, Eq.~(\ref{operator}). Indeed, let the
three-dimensional vector of displacement, $u$, be a field. Two equations for the
longitudinal  and transverse  elastic
waves  have the same form \cite{Landau-elastV7},
\begin{equation}
\frac{1}{v^2} \frac{\partial^2}{\partial t^2} u(x,t) =  \triangle u(x,t).  \label{elastic}
\end{equation}
The velocities of propagation of these waves are different and can be also
expressed through the  elasticity parameters \cite{Landau-elastV7}, which show
that longitudinal velocities are always greater than transverse ones. The
sound velocities can be also expressed through the elements of the elastic
stiffness tensor (the elastic constants or moduli), $c_{ij}$, and the matter
density, $\rho$, \cite{Royer-book2000}. There are, in general, three different
velocities (one longitudinal and two transverse) for each principal
crystallographic direction, for the cubic crystal lattices, which we consider as a simple example of this formalism. Due to the cubic lattice symmetries, there are only three independent elastic constants, $c_{11}, c_{12}, c_{44}$ \cite{Royer-book2000}. They determine velocities of the plane waves incident to the surface \{100\} (we use the notations of \cite{McSkimin-JAP1963}), 
\begin{equation}
v_1= \Big(\frac{c_{11}}{\rho} \Big)^{1/2}, \ v_2= \Big(\frac{c_{44}}{\rho} \Big)^{1/2},
\label{100}
\end{equation}
to the surface \{110\},
\begin{equation}
v_3= \Big(\frac{c_{11}+c_{12}+2  c_{44}}{2\rho} \Big)^{1/2}, \ 
v_4= \Big(\frac{c_{44}}{\rho}\Big)^{1/2}, \ 
v_5= \Big(\frac{c_{11} - c_{12}}{2\rho}\Big)^{1/2}.
\label{110}
\end{equation}
For the transverse wave $v_2$ both polarizations have the same velocity. We accept the elasticity theory's assumption that sound waves with different velocities are independent
\cite{Kittel-book1996} and, therefore, they present independent contributions to the
free energy. One can find sound velocities from the known elastic moduli
$c_{ij}$, but in practice the velocities are experimentally determined using the
ultrasound techniques, and the elastic moduli are derived. These high precision
measurements were made by H.J. McSkimin and P. Andreatch Jr. for some relevant
solids: diamond \cite{McSkimin-JAP1972}, silicon \cite{McSkimin-JAP1964},
germanium \cite{McSkimin-JAP1963}, gallium arsenide \cite{McSkimin-JAP1967} (where all the values are tabulated).   The third-order elastic moduli were also measured \cite{McSkimin-JAP1967}, they introduce nonlinearity to the wave equations.

The solutions for standing sound waves are obtained via the wave
equations \cite{Vladimirov-book1971}. This is done with help of the
expansion over modes that are counted by the frequency, a variable
reciprocal to time, but in the context of our problem the wavelength is an
appropriate variable. However, in a physical theory of heat we do  not  really need
actual solutions of differential equations. We want to obtain the total
energy of sound waves as a function of temperature, i.e., to find a scalar
function of one argument (with fixed geometrical charactristics of a body),
then the time (or frequency) is a redundant variable. The heat kernel method
allows us to obtain the trace of the heat kernel, which serves as a
mathematical prototype for free energy, directly, by skipping explicit solutions
of the field (wave) equations. Thus, usual derivations of the heat trace
asymptotics from the Laplacian's discrete spectrum can be avoided, if we seek
solving physics problems in the field theory. 

Since this computation is done in spacetime with the imaginary time, we first
go from the wave equations (\ref{elastic}) to the equivalent Laplace
equations. The spacetime metric changes its signature from Lorentzian to
Euclidean, which means that the l.h.s. of Eqs.~(\ref{elastic}) gains the minus
sign. In QFT literature \cite{DeWitt-book2003,GAV-lectures2007}, this is
commonly referred to as the Wick rotation. Thereby, we now have a Laplace
equation with the {\em four-dimensional} operator,
\begin{equation}
\Box =  \frac{1}{v^2}\frac{\partial^2}{\partial t^2} + \triangle .
\label{e-operator}
\end{equation}
Then, instead of solving the Laplace equation we solve the corresponding heat
equation (\ref{heateq}). This way we reduce a mathematical problem of the Debye
theory to the standard problem of the Schwinger-DeWitt QFT. Its operator
(\ref{operator}) accepts  the simplest form, where are no internal degrees of
freedom (no matrix structure \cite{BarVilk-PRep1985}), no potential term,
and no connection in the covariant derivatives: these are just
partial derivatives. The Laplace operator (\ref{e-operator}) allows
the separation of variables and thus the computation of the heat
kernel as the standard thermal sum (\ref{TrKbeta}). 

Finite temperature QFT, modified for condensed matter physics in
Sect.~{\ref{QFT}}, gives the volume contribution (\ref{Fba}) for each
independent velocity. The boundary contribution is defined by
its own surface wave velocity, $v_{s}$.  The total free energy is a
sum of all these terms,
\begin{equation}
F_{\mathrm{total}} = \sum_{i} F^{\alpha_i}[{\mathcal{V}}]  +
F^{\alpha_{s}}[{\mathcal{S}}]. \label{FEtotal}
\end{equation}  
For cubic lattices, the sum spans over nine $\alpha_i$, as the transverse wave
contributions from the waves incident to the surfaces \{100\} and \{111\}
double. The boundary contribution of the surface elastic waves, the Love or Rayleigh
waves \cite{Love-book1944}, may be neglected so far because it is an order of
magnitude smaller \cite{Gusev-FTQFT2015} in the experiments we consider below. 

The significant restriction is that sound velocities should be independent
of temperature. This condition was experimentally examined for the carbon
group elements \cite{McSkimin-JAP1972,McSkimin-JAP1964,McSkimin-JAP1963}. 
For example, it was found  that the sound velocities in diamond in the
temperature range, $-195.8{}^{\circ}C$ to $+50{}^{\circ}C$, change relatively as
little as $10^{-5}$. This is a good confirmation of the condition
{$v(T)=$const}. Further experiments with germanium and silicon were done \cite{Hao-PRB2001} in the temperature range, $25K<T<70K$, and the relative changes in
the longitudinal sound velocities were found to be about $10^{-4}$, with
germanium showing the higher change. Experiments at temperatures beyond this
range, which are important for the low temperature regime, were not done.
Nevertheless, the obtained results gives us hope that the relative
changes of velocities beyond the investigated range are also small. 

The phase velocity of plane waves are used in the elasticity theory derivations
\cite{Kittel-book1996,de-Launay-book1956}. In the heat kernel method, the {\em measured}
velocity enters the Laplacian (\ref{operator}) and correspondingly the $\alpha$-variable, (\ref{alpha}). Therefore, it is the velocity of energy transfer, i.e., the group velocity. This fact emphasizes the phenomenological nature of the proposed theory. There are many more issues with the use of elastic waves in thermal physics \cite{Kittel-book1996} that require
further development of the basic ideas outlined here.

\subsection{Molar specific heat}  \label{specificheat}

In Sect.~\ref{QFT}, we derived the dimensionless derivative (\ref{derivative}),
let us now propose the {\em scaling hypothesis}: scaling of a physical
observable, e.g., heat capacity, with the change of a physical variable, e.g.,
thermodynamic temperature, is the same as scaling of the corresponding
dimensionless functional, e.g., $\partial F^{\alpha}/\partial \alpha$, with the
change of a dimensionless variable, when it is expressed with help of the
fundamental physical constants.  It is the postulate that will allow us to avoid
solving the measurement problem.

Therefore, in order to obtain a physical observable, with the proper physical dimensionality, $J K^{-1}$, from the derived mathematical expression (\ref{derivative}) we multiply it with the Boltzmann's constant,
\begin{equation}	   
	C_V \equiv k_{\mathsf{B}} \frac{\partial F^{\alpha}}{\partial \alpha}, \label{CV}
\end{equation}
where $C_V$ is the heat capacity at the constant volume. We get,
\begin{equation}	   
	C_V =
\tilde{A} k_{\mathsf{B}} \frac{3}{\pi^2}  \frac{\mathcal{V}}{\tilde{a}^3} \, 
	\Theta(\alpha), \label{CVA}
\end{equation}
where volume $\mathcal{V}$ is fixed by the used mathematics, Sect.~\ref{QFT}. However, in experiments the heat capacity should be measured at fixed pressure to avoid changing the elastic properties of the crystal due to an additional strain \cite{McSkimin-JAP1963}. Thus, we take the experimental values for $C_P$. It is the heat capacity {\em per mole} that is derived here any way.  We do not treat the thermal expansion here, which could probably account for the higher temperature behaviour close to the critical points. 

In order to translate the expression (\ref{CV}) to the normally used molar
(atomic) specific heat, $C_M$, we should divide $C_V$ by the amount of
substance, $n$ (mol), contained in the given volume. The molar quantity, $n$,
can be found, with help of the Avogadro constant, $N_A$, if we know the number
of atoms $N$, as $n=N/N_A$. The number $N$ can be determined from
the system's volume, if know the volume of a lattice unit cell, $V$, and the
number of atoms per cell, $m$, as $N=m \mathcal{V}/V$. The lattice unit cell's
volume is defined by the lattice constants that correspond to a crystallographic lattice. For cubic lattices \cite{Kittel-book1996,Szwacki-book2010,Royer-book2000}, there is a single lattice constant, $a$, so, $V=a^3$. Finally, the amount of substance is,
\begin{equation}
n = \frac{\mathcal{V} m}{a^3 N_{\mathsf{A}}}. \label{moles}
\end{equation}
When Eq.~(\ref{CV}) is divided by (\ref{moles}), the molar specific heat (or
just the specific heat) becomes,
\begin{equation}
C_M = C_V \frac{a^3 N_A}{\mathcal{V} m} = 
\tilde{A} k_{\mathsf{B}} N_{\mathsf{A}} 
\frac{3}{\pi^2}  \frac{a^3}{m {\tilde{a}}^3}\, 
	\Theta(\alpha). \label{CMA}
\end{equation}

This expression still depends on the unknown cut-off constant, $\tilde{a}$.
However, $\tilde{a}$ is the length, which defines the limit of
validity of the elastic model, so it can be declared proportional to
the lattice constant,
\begin{equation}
\tilde{a} = B a.
\end{equation}
This gives us another calibration constant, $B$. As mentioned, we could
introduce it into the Planck's inverse temperature (\ref{beta}), but because
temperature comes only as the combination (\ref{alpha}), it does not matter if
we assign the uncertainty to $\tilde{a}$. The overall combination of the
calibration constants can be denoted, $A \equiv \tilde{A}/{B^3}$. Thus, as
expected, we arrive at the regulator-free quantity,
\begin{equation}
C_M =  A k_{\mathsf{B}} N_{\mathsf{A}} 
\frac{3}{\pi^2} \frac{1}{m}\, 
	\Theta(\alpha), \label{CMfinal}
\end{equation}
with
\begin{equation}
\alpha= \frac{\hbar v}{B k_{\mathsf{B}} a T }, \label{alphaB}
\end{equation}
which should be calibrated by experimental data in order to determine
the parameters $A$ and $B$. These two calibration constants define  scaling
of the specific heat, $C_M$. The constant $A$ sets up a vertical scale, while
the constant $B$ scales up $C_M$ horizontally.  

The formula for the specific heat (\ref{CMfinal}) is dimensionless, except for the factor of 
the molar gas constant, which is the only combination of fundamental
physical constants, $R= k_{\mathsf{B}} N_{\mathsf{A}}$, that could make up the
right dimensionality of specific heat. The expression for $C_M$ depends on the
velocities of sound, $v$, and temperature, $T$, that enter (\ref{alphaB}). The
sound velocities, in turn, are defined by the elastic constants and by the
matter density. So, despite its simplicity, Eq.~(\ref{CMfinal}) embodies  all
elastic characteristics of crystalline bodies, yet, its functional behaviour is
described by the universal thermal sum $\Theta(\alpha)$, Eq.~(\ref{Theta}) of
the single variable, $\alpha$.

The 'low temperature' asymptotics of the $\alpha$-derivative (\ref{lowalpha})
generates the term, cf., \cite{Gusev-FTQFT2015}, 
\begin{equation}	   
	C_M = 
	A k_{\mathsf{B}} N_{\mathsf{A}}
	\frac{\pi^2}{30}\, \frac{1}{m}\,
 \frac{1}{\alpha^4}, \ \alpha \rightarrow  \infty , \label{CvlowT}
\end{equation}
because it is a limit of the vanishing lattice constant, i.e., ideal (smooth)
medium. The 'high temperature' asymptotics (\ref{highalpha}) provides, via the
$\Theta(\alpha)$ maximum value (\ref{Thetamax}), the constant term independent
of $\alpha$,
\begin{equation}	   
	C_M = A  k_{\mathsf{B}} N_{\mathsf{A}}
	\Theta_{\mathrm{max}} \frac{3}{\pi^2}  \frac{1}{m}, \ \alpha 
\rightarrow 0. \label{CvhighT}
\end{equation}
The above expressions (\ref{CMfinal})-(\ref{CvhighT}) are valid for cubic
lattices, whose velocities are (\ref{100})-(\ref{110}). For other types
of lattices the final derivations should be redone starting from
Eq.~(\ref{moles}). This asymptotics can be also written in the form,
\begin{equation}	   
	C_M = \mathcal{R}
	\Theta_{\mathrm{max}}, \ \alpha  \rightarrow 0. \label{Rsolid}
\end{equation}
In this form, it lets us obtain the solid state equivalent of the molar gas constant, $\mathcal{R}$, which is different for different lattices. This factor can be used in the main Eq.~(\ref{CMfinal}) instead of the complex combination.

Even with the known elastic moduli (or velocities) we still have to calibrate the theory.
But after the main equations of the theory is derived, in principle, we can  forget about
elasticity theory.  We could declare the variables $\alpha_i$ be the ratios of thermodynamic temperature and some characteristic temperatures, $T_{i}$,
\begin{equation}
	\alpha_i \equiv T_{i}/T.
\end{equation}
In such a form, the theory would be more similar to the Debye theory, Appendix
A. However, this procedure would introduce many calibrating parameters in place
of just one, $B$. 

\subsection{High temperature limit of specific heat, or On the Dulong-Petit law}  \label{dulong}

In 1819  French natural philosophers Pierre Louis Dulong and Alexis Th\'er\`ese
Petit published  \cite{Dulong-ACP1819} the discovery that now bears their names.
A.T. Petit and P.L. Dulong measured how fast solid bodies cooled down in low
pressure air and found that the specific heat capacities of the bodies were
inversely proportional to the atomic weights of the bodies' chemical elements
\cite{Kittel-book1996,Laing-JCE2006}. Therefore, the Dulong-Petit law states
that the product of the molar mass, $M$, with the specific heat, $C_V$, of
mono-atomic solid bodies, i.e., the molar specific heat, $C_{M}$, is
approximately constant,
\begin{equation}
C_{M} = M C_{V} \approx 25 \  J/(mol \cdot K) 
\approx 6 \ cal/(mol \cdot K) \label{Dulong}. 
\end{equation}  
The Dulong-Petit law was important not only in physics, but also in chemistry, where it helped Dmitry Mendeleev to discover the periodic table of chemical elements \cite{Laing-JCE2006}. Mendeleev used the Dulong-Petit law to correct wrong atomic weights of three elements (cesium, uranium and indium) by making new measurements of their specific heats. 

Almost a century later, after the Avogadro constant, $N_{A}$, was introduced, Albert Einstein \cite{Einstein-AdP1907}  used  the Boltzmann's equipartition theorem to propose that the molar specific heat is universally equal to, 
\begin{equation}
3 R = 3 k_{B} N_{A}\approx  24.943 \  J mol^{-1} K^{-1}. \label{Rgas}
\end{equation}
He attempted to explain this empirical rule (\ref{Dulong}) as the classical limit
of a theory based on the new at that time Planck's quantum hypothesis. The
Dulong-Petit law in the form (\ref{Rgas}) was later re-derived in the Debye
theory \cite{Debye-AdP1912}, and it eventually became a dogma of solid state
physics \cite{Kittel-book1996} and thermochemistry \cite{Stoffel-ACIE2010}. 
However, this law does not hold strictly, and the Dulong-Petit (DP) values vary
widely \cite{Kittel-book1996}. In many cases such variations arise from
insufficient experimental data, as room temperature, assumed to be the high
temperature regime for the Dulong-Petit law, is certainly not high enough for
most substances. As is seen from the comprehensive analysis of numerous high
temperature experiments specific heats, e.g., for iron and silicon \cite{Desai-FeSi1986}, do not becomes constant, but keep slowly increasing to the melting point. Thus, the melting (or, in general, critical) points should provide the DP values of specific heats.

The $\alpha \rightarrow 0$ asymptotics  (\ref{CvhighT}) derived in this field theory formalism  corresponds to the Dulong-Petit limit.  In finite temperature QFT for condensed matter, the Dulong-Petit limit emerges from the minimum length proportional to the lattice constants that cuts off the range of short wavelengths, when the field theory ceases to be
valid. In other words, the approximation of condensed matter as continuous medium fails, when the elastic wavelengths become comparable to the average distance between the nodes of lattice.

The Dulong-Petit limit is a constant independent of temperature and elastic properties of the lattice, but not of its crystallographic type. This high temperature limit depends
on the number of points in a unit cell, $m$, and the relative volume of a unit
cell (one for cubic lattices). It should also depend on the total number of
velocities of a lattice contributing to free energy (\ref{FEtotal}). This means
{\em the Dulong-Petit limit should be the same for all substances that crystallize to
the same lattice type}.

The Avogadro constant and the Boltzmann's constant enter
the Dulong-Petit limit, because it is the only possible combination of
fundamental physical constants with the right dimensionality for specific heat.
Therefore, we argue that proportionality of the Dulong-Petit limit
(\ref{Dulong}) to the molar gas constant (\ref{Rgas}) follows trivially  from
dimensional analysis, while  {\em its numerical factor close to 3 is merely a
coincidence}. The number three has never been well measured, and its appearance
seems to be due to the magic of whole numbers that influenced theoretical
developments.  All the reasonings proposed as its proofs are based on a
discrete model of independent oscillators
\cite{Kittel-book1996,Einstein-AdP1907,Debye-AdP1912}, which apparently failed to correctly describe specific heat for other temperature ranges. However, we could expect that statistical physics, based on new mathematics, being developed by Victor Maslov \cite{Maslov-MN2013} should produce a correct version of the discrete theory for the lattice specific heat.

\subsection{Finite size effects} \label{size}

Free energy in the finite temperature QFT   based on the heat kernel (\ref{TrKlocal}) contains two terms defined by geometrical invariants of a compact domain, the volume and the area of boundary \cite{Gusev-FTQFT2015}. We focus in this work only on the volume ('bulk') term as it gives the leading order contribution in experiments we want to consider. Let us sketch the reason for discarding the boundary term.

With a system's effective size \cite{Gusev-FTQFT2015}, defined as the ratio of the body's volume to its boundary's area, $ r = {\mathcal{V}}/{\mathcal{S}}$, the acoustic kappa-factor is determined by the sound velocity, $v\propto 10^3\, m/s$,
\begin{equation}
\kappa_a \propto {\hbar v}/{k_{\mathsf{B}}} \approx 1.1 \cdot 10^{-9}\ K \, m.
\end{equation}
It shows that  threshold for the appearance of the boundary (finite size) effects in the
lattice specific heat is many orders of magnitude higher than in thermal
radiation phenomena, where $\kappa \approx 3.3 \cdot 10^{-4} K\, m$,
\cite{Gusev-RJMP2014}. Assuming an experimental uncertainty is $1\%$, we
expect that the finite size effects at low temperatures, e.g., $1< T<10\, K$, could appear only in systems of a nanometer size, i.e., with $r\propto  10^{-9}\,m$. In the specific heat experiments with 'macroscopic' samples, \cite{Desnoyers-PhilMag1958,Flubacher-PM1959,Berg-PRSA1957}, the finite size effects can be seen only at sub-Kelvin temperatures.  In general, the true low temperature asymptotics, $\beta \gg   r$, depends on the shape and the surface curvatures of a condensed matter system, as well as on its material. Therefore, a universal low temperature asymptotics of the free energy (and of the specific heat) does not exist. Detailed analysis of finite size effects in the heat capacity certainly deserves a special study motivated by the recent experimental quest in nanotechnology.

\subsection{Specific heat of compounds, or On the Neumann-Kopp law} \label{NKrule}

The properties of specific heats of compounds (multi-atomic substances) have been studied  in
the 19th century by German physicist  Franz Ernst Neumann and German chemist Hermann Franz Moritz Kopp. They  formulated the rule (their independent works separated by a time interval) for specific heat of a chemical compound as equal to the sum of specific heats of its constituents. As every empirical law, the Neumann-Kopp  rule does not hold strictly.  When this rule was derived ambient conditions were considered to be sufficient to reach the Dolung-Petit limit, but they are not. The vast number of new compounds were discovered and studied. Extensive data of modern measurements show that many compounds do not obey the Neumann-Kopp rule, for example, the overview of solid binary antimonides \cite{Schlesinger-ChemRev2013} states that {\em ''the Neumann-Kopp law is not a suitable substitute for experimentally obtained values of antimonide heat capacities''}. Likely this is the case for other classes of compounds.

In general, the Neumann-Kopp rule would give n-times bigger a specific heat for an n-atomic compound, only if the Dulong-Petit law were exact and hold at 'ambient conditions'; this is not the case. Let us have a fresh look at this rule from the point of view of the geometrical formalism, where the heat capacity of a crystal is defined by the number of sound velocities and lattice characteristics. This means that specific heats of elements that make up a compound should be irrelevant. 

The Neumann-Kopp rule for specific heats of crystal matter has been established, because it uses the definition of the amount of substance (mole) of solid state matter derived for gaseous matter. Thus, one mole of an n-atomic compound contains n-times more constituents (atoms as the lattice nodes) than all its composing elements in solid state contain. If the composing elements crystallize to the same type of lattice as the compound, then the Neumann-Kopp rule would be exact. This is, of course, a special case, but many common substances have similar specific heats under some commonly used conditions (close to the Dulong-Petit limit). It is clear that for many more substances, which were discovered and studied after the 19th century, under general conditions, the Neumann-Kopp rule is false.

Therefore, we suggest that one mole of crystal matter should {\em not} be equal to $N_A$ {\em molecules}, but rather to the Avogadro number of atoms (or ions) as true constituents of a crystal lattice. Indeed, separate molecules of a chemical compound do not exist in a crystal. Instead, the whole crystal lattice can be viewed as one large molecule of a compound, whose atoms take places at the lattice nodes.  We propose to correct tabulated specific heats for n-atomic compounds by dividing their values by n. As an example, we will consider two-atomic compounds with the {\em zinc-blende} lattice, since it is a cubic lattice of the diamond type for two-atomic materials. We argue that the corrected Dulong-Petit limit of such substances is equal to the one for the carbon group elements with the diamond lattice.

\section{Experiments. Specific heat of the diamond lattice}
\label{experiment}

\subsection{The group IV elements}

In order to examine the presented theoretical ideas, let us consider
experimental determinations of specific heats of some solid substances.  
We will do the statistical analysis of available data for the elements from
group IV (carbon group) of the periodic table of chemical elements that
crystallize to the diamond type cubic lattice. They include diamond (carbon, C),
silicon (Si), germanium (Ge), and gray tin ($\alpha$-Sn). These are
$\alpha$-forms (diamond lattice) of Si and Ge, but  their $\beta$-forms with
tetragonal lattices exist under certain pressures \cite{Greenwood-book1998}. 
The properties of gray tin are little measured, so we try to make some
predictions.

In 1950s, J.A. Morrison with collaborators performed at chemical laboratories of
the National Research Council of Canada  precise measurements of specific heats
of many crystalline solids at a wide range of temperatures. These were traditional
measurements done with the calorimeters. For some of these solids, the
Morrison's experiments remain a state of the art today, because no attempts were
made to repeat them with higher precision. Morrison's group measured specific
heats of natural diamonds \cite{Desnoyers-PhilMag1958} and pure, commercially
grown, single crystals of silicon and germanium \cite{Flubacher-PM1959}. They
published full tables of the original experimental data  together with the
analysis. Among the three data sets available for the carbon group
elements, the data for germanium and silicon \cite{Flubacher-PM1959} are the best. Likely this is because the diamonds used in the study were natural, thus, with uncontrolled defects and impurities. We use the germanium data as an instance of the diamond lattice although the silicon data are as good and produce the same physical results. All measurement data were converted to the SI units $J\, mol^{-1}\, K^{-1}$. 

The number of atoms per unit cell, $m$, of the diamond lattice is eight \cite{Kittel-book1996}. The velocities of sound in crystals of the carbon group elements were measured in Refs.~\cite{McSkimin-JAP1972,McSkimin-JAP1964,McSkimin-JAP1963}. The derived elastic constants are given according to these papers in the units of $GPa$, in Table~\ref{group4}, as a more concise and equivalent representation. The elastic constants of gray tin were implied from the neutron scattering experiments \cite{Price-PRB1971}, the standard technique to determine frequencies of crystals; these values are accepted in the later references like \cite{Adachi-book2005}. The lattice constants, $a$, in \AA nsgtrom, $10^{-10}\, m$, are from the reference book \cite{Adachi-book2005}. The densities, $\rho$, also from \cite{Adachi-book2005}, are expressed in the commonly used units of $g/cm^3$. The lowest velocities of sound, $v_5$, in units of $10^3\, m/s$, are also given (it is computed for $\alpha$-tin). Critical temperatures, $T_{\mathrm{cr}}$, are the sublimation temperature for diamond \cite{Pierson-book1993}, the melting points of Si and Ge \cite{Adachi-book2005}, and the $\alpha \rightarrow \beta$ lattice transition (its reverse is the 'tin pest' effect \cite{Greenwood-book1998}) for $\alpha$-tin. The low-$T$ characteristic temperature, $T_0$, is explained below.  The values not directly measured yet (predicted) are denoted with the asterisk sign. The elastic moduli for GaAs are from Ref.~\cite{McSkimin-JAP1967}. Its highest specific heat, given in Ref.~\cite{Glazov-IM2000}, is corrected by the factor 1/2. Some of the basic properties of these and other semiconductors are also available online at Ioffe Institute \cite{IoffeInst-2001}.
\begin{table}[!ht]
\caption{Properties of the carbon group elements and GaAs}
\label{group4}
\begin{tabular}{lllllllllll}
\hline
material& a & $c_{11}$&$c_{12}$& $c_{44}$&$\rho$& $v_5$ & $T_0$&  $\theta$ &$T_{\mathrm{cr}}$& $C_{\mathrm{DP}}$\\
\hline
diamond    & 3.567 & 1079 & 124.0 & 576.0  & 3.5156 & 11.659 &173.3 & 1.44 &3900 & 27.5 \\
$\alpha$-Si& 5.431 & 165.78 & 63.937 & 79.625 &3.3291 & 4.6739 & 39.4 & 1.67 &1685 & 29.16 \\
$\alpha$-Ge& 5.657 & 128.53& 48.26 & 66.80  & 5.3256 & 2.7459 & 21.4 & 1.73 &1210 & 28.76 \\
$\alpha$-Sn& 6.4892&66.7&36.5 &30.2&5.7710& 1.618${}^{*}$&11.0${}^{*}$&1.73${}^{*}$&286 &29.1${}^{*}$\\
GaAs   & 5.6533& 118.77 & 53.72& 59.44& 5.3175 & 2.4732 & 20.0 & 1.67 & 1513 & 29.08${}^{*}$ \\
\hline
\end{tabular}
\end{table}

\subsection{The Dulong-Petit values} \label{DPvalues}

We need to know the high temperature limit (the Dulong-Petit value) of specific
heat of the carbon group elements with the diamond lattice (or any other elements with this lattice) for calibrating the model. At the same time we check the hypothesis that the
Dulong-Petit values of all such elements are the same as suggested by Eq.~(\ref{CvhighT}). 

The Dulong-Petit value for diamond is the most difficult to obtain, because its
$\Theta(\alpha)$ spreads wide, thus, its low temperature regime is high,
$T < 170K$. The only reliable number is $C_{\mathrm{DP}}=22.10\, J/(mol\,
K)$ at 1100K in Ref.~\cite{Victor-JCP1962}, but its temperature is too low for
the anomalous thermal behaviour of diamond. Two more numbers, 24.7
and 26.3 $J/(mol\, K)$, at correspondingly 1800 and 3000K, are given in
the reference book on carbon \cite{Pierson-book1993}, but their original source
is not clear. Since the monotonic growth of specific heats is almost linear, we
make an extrapolation to the critical (sublimation) temperature  3900K to
get $C_{\mathrm{DP}}\approx 27.5 , \ J/(mol\,K)$. The reliably measured DP limit of diamond is apparently not known.

The DP value for silicon, 29.199 J/(mol\, K) at its melting point  is
taken from the reference book \cite{Barin-book1995}. Ref.~\cite{Glazov-HT2001} cites the book \cite{Gurvich-book1990} for the recommended polynomial fit, which gives 29.114 J/(mol\, K). We take their average as $C_{\mathrm{DP}}$.

The specific heat  of germanium near the melting point is given in the reference book
\cite{Barin-book1995}: $C_{\mathrm{DP}} = 28.756 J /(mol\, K)$. It lets us fix the
overall constant of the model, (\ref{Rsolid}), for the diamond lattices: 
\begin{equation}
\mathcal{R}\approx 345\,  J\, mol^{-1}\, K^{-1}. \label{RGe}
\end{equation}

The Dulong-Petit value of gray tin could be measured at rather low
temperatures. Its critical temperature is marked by transition to $\beta$-Sn. We found
the $\alpha$-Sn specific heat $0.278\, J/(g\, K)$ at 100K in the reference book \cite{Adachi-book2005}, which translates to $33.00 \, J/(mol\, K)$. However, it is not measured, but implied from the elastic constants. It is clearly higher then other $C_{\mathrm{DP}}$ values in the table. According to the proposed model, we conjecture, that the true value of the Dulong-Petit limit of specific heat of gray tin is the same as the one for silicon and germanium, namely, approximately $29\, J /(mol\, K)$. 

Summarizing, it is likely, when experimental inconsistencies are cleared
up, the lattice specific heats of the group IV elements with the diamond lattice would be the same. This value is close to $29 J\, mol{}^{-1}\, K{}^{-1}$ and is certainly away from the presumed exact value (\ref{Rgas}). We will further check if this statement is true for {\em any} substance with the diamond lattice.

\subsection{Low temperature asymptotics}

\begin{figure}[h!]
  \caption{Finding the low temperature regime of Ge}
  \centering
\includegraphics[scale=0.5]{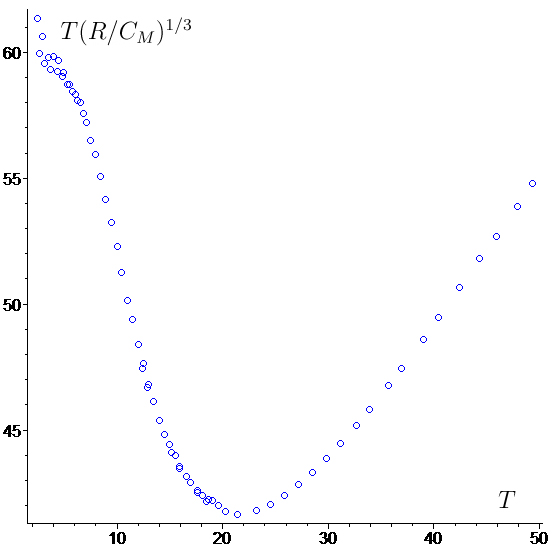}
\label{TC13}
\end{figure}

Let us explore the low temperature asymptotics of specific heats. It is common
in the literature \cite{Kittel-book1996} to plot the data not for specific heat,
but for the implied (effective) Debye temperature (\ref{DebyeT}). It is
experimentally known that the effective Debye temperature is not constant, as it
should be if the Debye theory were correct, instead it exhibits some complex
behaviour, especially at low temperature
\cite{Kittel-book1996,Gopal-book1966,Flubacher-PM1959,Berg-PRSA1957}. We use a
similar combination, $T(R/C_M)^{1/3}$, because it clearly distinguishes
different temperature regimes at low temperatures. The gas constant merely
allows us to have the units of Kelvin on both axes, while numerical values on
the vertical axis are irrelevant. The resulting graph for germanium in Fig.~\ref{TC13}
resembles typical curves for the Debye temperatures
\cite{Flubacher-PM1959,Berg-PRSA1957}. 
Because we propose that the low temperature behaviour of specific heat is $T^4$, we
interpret the descending branch of this graph as the power law $T^{-1/3}$.
Temperature at the minimum value of this graph, denoted as $T_0$ , indicates the
limit of the low temperature regime. It is the characteristic temperature for
any given material. For germanium, the characteristic temperature is $T_0=21.4\,
K$. Characteristic temperatures for diamond and silicon are given in
Table~\ref{group4}. The region of low temperatures for specific heat is limited
from below as well, because the relative contribution of the surface heat
grows in addition to the experimental uncertainty, which shows up in these
data.

Let us show that the characteristic temperature of a condensed matter system that specifies the condition $\alpha \gg 1$, also gives a universal heat capacity constant. Skipping the unessential calibration parameter, the explicit form of the low-$T$ asymptotics is,
\begin{equation}
\frac{\hbar v}{k_{\mathsf{B}} T a} \gg 1.  \label{lowTcond}
\end{equation}
We take the lowest velocity of sound, $v_5$, since it gives the leading contribution at low temperatures. The threshold for this condition to hold is given by the characteristic temperature, $T_0$. We conjecture that the l.h.s. of (\ref{lowTcond}), taken at this temperature, is a dimensionless constant, which is the same for all materials with the same type of crystal lattice,
\begin{equation}
	\theta= \frac{\hbar  v_5}{k_{\mathsf{B}} T_0 a}. \label{theta}
\end{equation}
The values of $\theta$ are given Table~\ref{group4}, which shows that they are indeed very close, with the diamond's value somewhat apart. Let us take now the computed sound velocity, $v_5$, for $\alpha$-tin to derive $T_0$ for gray tin from this constant (\ref{theta}), taken to be 1.73:
\begin{equation}
{\alpha{\mathrm{-Sn}}:}\	T_0 = \frac{\hbar  v_5}{k_{\mathsf{B}} \theta a}
\approx 11.0 \, K. \label{T0Sn}
\end{equation}
This is a testable {\em  prediction} for future determinations of the specific heat of single crystals of gray tin.

Let us now test the $T^4$-hypothesis (\ref{CvlowT}) vs. the Debye cubic law (\ref{DebyeLT})
statistically and graphically, without calibrating the full theory, as we need this for the second calibration parameter. We take the obtained low temperature range, $T=4.364\ldots 20.233\,K$, and fit its specific heat data with two testing functions, $aT^4$ and $bT^3$. The resulting least-square fits obtained with Maple are $a= 6.297\cdot 10^{-6}\ J/(mol K^5)$ and $b= 9.085\cdot 10^{-5}\ J/(mol K^4)$. The chi-square statistics for both hypotheses do {\em
not} reject either, i.e., both laws are statistically allowed. However, the $\chi^2$-statistic for the cubic law is 0.162 vs. 0.002 for the quartic  function, with both p-values equal to one; that shows the $T^4$-function is favored.

\begin{figure}[h!]
  \caption{Testing the low-T hypotheses for Ge. I}
  \centering
\includegraphics[scale=0.5]{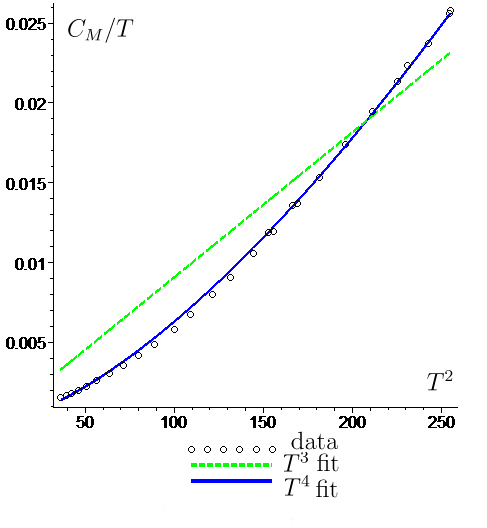}
\label{CvTT2}
\end{figure}

The more apparent evidence can be seen, when two obtained fits are plotted together with the measured data, in a popular graph, $C_M/T$ vs. $T^2$, Fig.~\ref{CvTT2}, often used to demonstrate validity of the Debye cubic law \cite{Kittel-book1996,Flubacher-PM1959,Berg-PRSA1957},
It is obvious that in the selected temperature range the specific heat of germanium does {\em not} obey the cubic law. 

\begin{figure}[h!]
  \caption{Testing the low-T hypotheses for Ge. II}
  \centering
\includegraphics[scale=0.5]{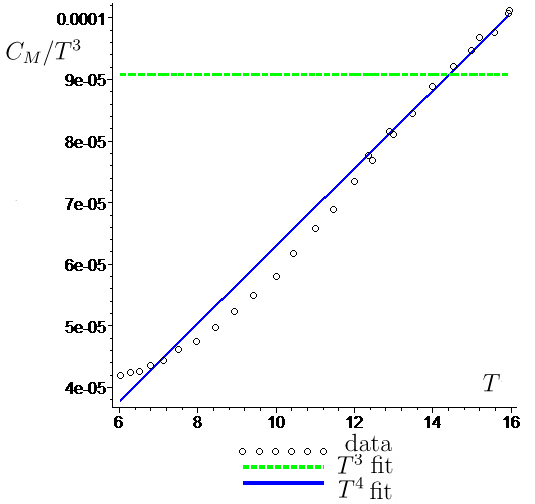}
\label{CvT3T}
\end{figure}

Another popular graph in the solid state physics literature is $C_M/T^3$ vs. $T$, Fig.~\ref{CvT3T}. 
This graph looks even more compelling, as the observed data are nowhere close to a constant (horizontal line) expected from the Debye $T^3$-law. Instead they can be approximated by a linear function (obtained above by fitting the original data, not this scaled data), which signifies a $T^4$-law. Note, these fits assumption, $C_M=0$ at $T=0$,  which is excluded in the geometrical formalism, is only a reasonable approximation here, as $C_M$ is infinitesimal (but never zero due to the surface heat) at $T < 1\, K$. 

The results of the power law fits and the $\chi^2$-statistics as well as the graphs made for the silicon data \cite{Flubacher-PM1959} are qualitatively the same, they confirm the quartic power law. 

The diamond data \cite{Desnoyers-PhilMag1958} gave inconclusive results, both the power laws are equal at approximating the low-T specific heat. The reasons could be not only the anomalously big region of low temperatures, but also the presence of impurities and defects in natural diamonds, or catalytic atoms in synthetic diamonds (also see Discussion).

\begin{figure}[h!]
  \caption{Testing the low-T hypotheses for Ge. III}
  \centering
\includegraphics[scale=0.5]{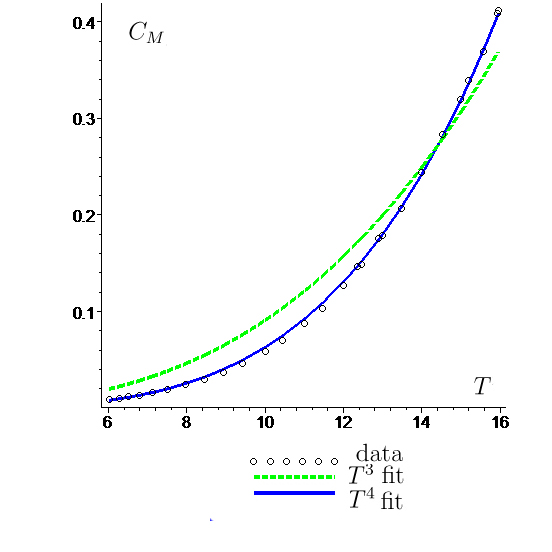}
\label{rawdata}
\end{figure}

In general, it is difficult to make a selection of the two tested power laws using original (not scaled) data. However, this is not in the case for germanium as seen from Fig.~\ref{rawdata}.
However, in the solid state physics literature, such fits  are done  not to the cubic law, but to the extended polynomial, $C_M= a\, T^3+b\, T^5 +c\, T^7$. Various arguments were given for its introduction, but its two higher power coefficients usually have low or no statistical significance.  Probably because of the use of such extended polynomials and because of the firm belief in the Debye cubic law, its validity has not been questioned, even though  experimental data contradicting to it are abundant. 

\subsection{Specific heat of gallium arsenide}

The ideas about the Kopp-Neumann rule for multi-atomic substances suppose that compounds have the same properties as elements if their crystal lattice types are the same. In the diamond lattice of a two-atomic compound (zinc-blende lattice), half of its atoms belong to each composing element \cite{Kittel-book1996,Szwacki-book2010}. Many semiconductors crystallize to the zinc-blende lattices \cite{IoffeInst-2001}, thus, their properties are well measured \cite{Adachi-book1992}.

Let us take as gallium arsenide, GaAs, an example.   According to the conjecture of Sect.~\ref{NKrule}, the Dulong-Petit limit of GaAs should be the same as the DP value of any of the carbon group elements with the diamond lattice, if we would take one mole of its {\em atoms}, not molecules. Indeed, this is true within experimental uncertainty: the GaAs specific heat given at $1500\, K$ in \cite{Glazov-IM2000} is $C_{\mathrm{DP}}=58.17/2 = 29.08 \ J/(mol
K)$.
 
We can also look at the low temperature asymptotics of the GaAs specific heat. The study \cite{Cetas-PR1968} provides the sufficient number of data points for specific heat of gallium arsenide. Using it we made a plot of the type of Fig.~\ref{TC13} that determined the low temperature region by $T_0=20.0\, K$, which produced the $\theta$-constant as 1.67, confirming its universal nature. Further statistical and graphical analysis performed according to the described above procedures gave the same conclusions as for the group IV elements: {\em the low temperature asymptotics of the GaAs  specific heat is a $T^4$-function}.

Therefore, we can conclude that the conjecture about similarity of thermal characteristics of compounds and elements with the same lattice type holds. In fact, their elastic properties are also similar as they should be. In particular, gallium arsenide resembles germanium, as is seen from Table~\ref{group4}. Such a similarity was noted first in \cite{McSkimin-JAP1967}, and it is likely the fundamental reason for a near coincidence of characteristics of the considered specific heats.

\section{Summary}

Chief applications of finite temperature quantum field theory lie in condensed matter physics. As a first element of this program we have implemented the idea of P. Debye about lattice heat as energy of sound waves in elastic bodies. The proposed thermodynamics of solid state matter presents the theory of specific heat formulated in the field theory language, without reference to the Planck's distribution and quantum oscillators. The summary is here.
\begin{itemize}
\item
Thermodynamics of condensed matter is a theory different from classical thermodynamics of gases.
\item
The dimensionless combination of thermodynamic temperature and the lattice
constant is a proper thermal variable in condensed matter physics.
\item
The dimensionless functional of free energy is defined by the sound
velocities, lattice constants, crystallographic type and geometrical
characteristics of a condensed matter system.
\item
The measurement postulate is expressed as the scaling hypothesis.
\item
Specific heat of any condensed matter system is determined by the universal
thermal sum of the new thermodynamic variable. 
\item
The specific heat near a critical point (the Dulong-Petit vale) is the same for all substances with the same type of crystal lattice. The Dulong-Petit limit appears due to the lattice cell size cut-off.
\item
One mole of an n-atomic compound, in a crystal form, should be equal to the Avogadro number of its atoms. The Neumann-Kopp law holds only in special cases due to the old definition of the solid matter mole.
\item
The lattice specific heat at low temperatures is the $T^4$ law, not the Debye
$T^3$ law. Experimental data of J.A. Morrison group and some other data decisively confirm the quartic law. Other power laws may appear due to the electronic, surface and
polycrystalline properties of solid matter.
\item
Universal thermal asymptotics of a solid body towards the absolute zero temperature do not exist, because they depend on the body's shape and material.
\end{itemize}

\section{Discussion}

The central object of spectral geometry \cite{Gilkey-book2004} is the trace of the heat kernel. Its leading (under the 'high-temperature' conditions \cite{Gusev-FTQFT2015}) terms are expressed by two geometrical invariants of a system's domain, volume and the boundary's area, (\ref{TrKlocal}). This theorem of mathematical physics has been re-derived by various methods starting from the pioneering works of P. Debye \cite{Debye-AdP1912} and H. Weyl \cite{Weyl-RCMP1915}. The heat kernel removes the need to study actual spectra of condensed matter systems. The spectrum of lattice frequencies, as studied in the neutron scattering experiments and in the phonon theory, may be complex, nevertheless, its heat kernel trace is always proportional to the system's volume, (\ref{TrKlocal}).

We need to know the function's values at two temperature points in order to finally fix it for a specific class of materials, e.g., diamond lattice crystals. Thus, the model should be calibrated anew for every type of crystal lattices, which means there is its own thermodynamics for every crystallographic type. This is not surprising at all. Classic thermodynamics was developed as the physics of rarefied gases, and the universal feature of gas is its homogeneity. In contrast, solid state matter is inhomogeneous and anisotropic, i.e., it possesses internal scales along different directions. These geometrical properties determine the thermodynamics of condensed matter systems.

We have not completed the full calibration of this model. In the low temperature regime, the model has a simple power behaviour, which is defined by the lowest velocity of sound. However, as seen from the $\Theta (\alpha)$ behaviour, Fig.~\ref{thetaalpha}, the one-velocity model can only be correct in the low temperature limit.  According to the assumption about the total free energy (\ref{FEtotal}), the sum over all allowed velocities should be performed. However, the naive sum without independent weights, i.e., with equal weights, $\mathcal{R}/9$, does not match the full data set. Obviously, by varying weights in the sum we could fit the data well, but this is an artificial procedure that would multiply the number of calibration parameters. This means we have to acquire more information about physics of elastic waves to finish this work.

Experimental testing of this model has only begun. The data for  natural \cite{Desnoyers-PhilMag1958} and synthetic \cite{Atake-RRLEM1991} diamonds do not confirm the quartic law. Instead both power laws are statistically allowed, but they are equally bad at describing the natural diamonds data, while the Debye cubic law is selected for the synthetic
diamonds. However, diamond is the single thermophysical anomaly among all
substances in the Nature; more precise measurements are needed. Furthermore, 
available data sets for metals, copper, Cu \cite{Cetas-PR1968} and aluminum,
Al \cite{Berg-PhysRev1968}, show the $T^3$ behaviors at low temperatures when
treated with the algorithms above. Whether this is a property of metals with the
electronic heat contribution or the characteristics of polycrystalline solids,
which apparently formed the used ingots, should be examined. At the same time we can observe the $T^4$-power law in the specific heat data of some two-atomic compounds: gallium antimonide, GaSb \cite{Cetas-PR1968}, silver chloride, AgCl \cite{Berg-PRB1976}. The critical experimental analysis and the theory development should continue.

Physical properties of polycrystalline bodies are different from those of single crystals. A polycrystalline body can be considered as a composite of grains  (domains) of single ideal crystals joined by their common faces (boundaries). Therefore, it is reasonable to assume that each grain has its own volume free energy as well as its own surface free energy, while the contributions from the edges and vertices of the grains could be significant. We know that the contribution of a boundary may exceed the volume one in the 'low temperature' regime \cite{Gusev-FTQFT2015}. Thus, the total heat capacity of polycrystalline bodies may be complex.

In his work \cite{Einstein-AdP1907} A. Einstein introduced the quantum hypothesis of M. Planck to condensed matter theory. It has been admitted that this innovative approach was not correct in the method and the result. The Einstein theory seemingly produced the Dulong-Petit law, but not the low temperature asymptotics. We have argued that the DP limit could also not be derived this way. The non-success was caused by the use of mechanistic methods,
inherited through statistical mechanics (oscillators and the equipartition theorem), in an entirely new realm of physics. In this early work Einstein did not get rid of classical mechanics altogether, but later he criticized {\em a mechanical view} of physics and advocated dispensing it later \cite{Einstein-book1967}. P. Debye chose the right approach, but the Planck's empirical formula \cite{Planck-book1914} remained the only way to connect elasticity theory and thermal physics. More critique of his theory is in Appendix A.

The  dynamical theory of crystal lattices developed for specific heat by Max Born and Theodore von Karman \cite{Kittel-book1996,Born-book1962} was originally close to the Debye theory \cite{Born-PZ1913} in its use of elasticity theory. In modern textbooks the Born-von Karman theory is considered the first principles theory based on quantum mechanics. The obvious and often criticized fault of this theory is its periodic (cyclic) boundary conditions (the Born-von Karman boundary conditions), which define a condensed matter system's space as compact without a boundary. Thus, this theory cannot deal with boundary effects at all. Furthermore, it is important to realize that topology of the Born-von Karman theory's space is non-trivial. Its boundary conditions effectively substitute topology of the open space, $\mathbb{R}^3$, with topology of the three-dimensional torus, $\mathbb{S}^1 \times \mathbb{S}^1  \times \mathbb{S}^1$. This makes it a different mathematical problem, whose solutions are also different from the original problem's ones.  The comparisons of the
Born theory's predictions with experimental data were done  and they demonstrated \cite{de-Launay-book1956} that the Born-von Karman theory had failed to agree with the available at that time experimental data. Later experimental studies  \cite{Victor-JCP1962} also cited significant disagreements of measured specific heats with theoretical results of the this theory. Therefore, we are forced to conclude that the Born-von Karman theory is false.

Another theory of the lattice specific heat is a phenomenological theory of
Chandrasekara Venkata Raman \cite{Raman-PIASA1956}. His theory of the crystal lattice dynamics  was based on the frequencies of a basic block of atoms, e.g., in the diamond lattice \cite{Raman-PIASA1956-2}, and it is apparently good at describing some properties of diamond's specific heat \cite{Victor-JCP1962}. However, Raman's theory is not working at low temperatures \cite{Victor-JCP1962}. The likely reason is that in this regime the theory should take into account the system's spatial domain as a whole, i.e., the full lattice with its boundary, and the discreteness of its frequency spectrum.

Dissatisfaction with the Debye theory brought up attempts to improve its agreement with experiment by introducing more parameters to fit the empirical equations to the observation data. Sometimes large discrepancies between the Debye theory and experiments are assigned to peculiar properties of a specific state of matter, such as glass \cite{Matsuda-JNCS2011}. However, the shape of the low-temperature graphs in Ref.~\cite{Matsuda-JNCS2011} looks similar to Fig.~\ref{CvT3T}. Thus, it could likely be a consequence of the universal power law (\ref{CvlowT}), which makes its appearance in many instances of condensed matter.

Diamond is one of the most experimentally  studied crystals. It has fascinated
people and researchers for a long time,
\cite{Einstein-AdP1907,Raman-PIASA1956,Desnoyers-PhilMag1958,Atake-RRLEM1991}.
Diamond is indeed a thermal anomaly among elements, and it proved its character again
by refusing to reveal its low temperature power law. The belief in the anomalously low Dulong-Petit limit of diamond \cite{Kittel-book1996} was false as its specific heat maximum is simply reached at very high temperatures. Diamond's remarkable thermal properties are explained by its unusual elastic characteristics. Namely, the high velocities of sound and the dense lattice provide the anomalously large combination $v/a$ in (\ref{alphaB}),
which spreads the functions $\Theta(\alpha)$ along the temperature axis. Thus, the diamond gets the longest range of low temperatures. We believe the parameters in Table~\ref{group4} found in experiments should be different if ideal diamond crystals were available for measurement. Neither natural, nor currently produced synthetic (due to the used technology) diamonds can be satisfactory. In particular, the DP value should be the same as for other elements, about 29.1 J/(mol K), and the $\theta$-parameter be equal to 1.7. The importance of carbon for the technology of high temperature resistant materials gives us a hope that these numbers will be verified soon.

It is derived in statistical mechanics that the specific heat and the heat conductivity are intrinsically related. Therefore, often instead of direct determinations, the heat conductivity is measured while the specific heat is implied from the Fourier's thermal conductivity  equation \cite{Kittel-book1996}. Since we have not derived yet thermal conductivity in the present formalism, we have focused only on the calorimeter measurements. For this reason, we
place the discussion of specific heat of solid argon \cite{Finegold-PR1969,Flubacher-PPS1961}, the textbook's topic, Chap. 5 of \cite{Kittel-book1996}, to Appendix B.

To a great surprise, basic thermal and elastic properties of $\alpha$-tin, in single crystals,  were not studied yet (at least we failed to find such studies and data). Nor specific heat, neither sound velocities were directly measured in gray tin. The technique to produce single crystals of $\alpha$-tin is described in Ref.~\cite{Styrkas-IM2005}. Now these crystals await an experimentalist to have their properties measured.

A special statement about experimental data is in order. It is regretful that
starting from around 1980s, many reports on physics experiments stopped
including essential measurement data in their publications. Instead, data are
analyzed by experimental teams, sometimes in collaboration with
theoreticians, and a result of such an analysis is presented. If data are
presented, it is done in plots, which makes them useless for other researchers.
This situation creates a serious danger to physics as an experimental
science by removing access to valuable information that should belong to the whole
science community. Appendix C of this paper reproduces the specific heat data
for germanium, Ref.~\cite{Flubacher-PM1959}, that were obtained at the
publicly funded laboratories of National Research Council of Canada. Let us 
mention that the results of another great work of J.A.  Morrison's group
\cite{Berg-PRSA1957} were essentially lost, as we failed to recover its full
data, deposited at the Royal Society (U.K.) archives, despite the efforts of
both countries' agencies.

When the present work was done and the paper was written, we discovered the comprehensive series of  works of Roland P\"assler, e.g., \cite{Passler-PSSB2008,Passler-PSSB2007}. In these works a new model for the description of specific heats over an entire range of measured temperatures is developed. The model contains several characteristic temperatures and fits the data with the sums of exponential functions, whose arguments are quadratic in the temperature. The fact that data of Ref.~\cite{Flubacher-PM1959} are used to calibrate the model further emphasizes the crucial role of experimental data for theory development. A careful comparison is required, but it is already clear that the empirical representation derived by P\"assler is similar in some aspects to the axiomatic proposal above.

Thermodynamics was created as physics of gaseous state of matter, and its main practical application was the development of heat engines. Therefore, in its original form classical thermodynamics is limited in the range of physical phenomena it can describe, and its extension to continuous medium requires further development. Thermodynamics is incomplete until a body's {\em boundaries} included into a physical theory. In classical thermodynamics, boundaries are present, but only implicitly as a means to keep volume finite and to transfer heat energy from gas to external media and reverse. Thermodynamics with boundaries included {\em explicitly} arises from the finite temperature quantum field theory. 

The opening paragraphs of the paper by James Clerk Maxwell 'On the dynamical
theory of gases' \cite{Maxwell-book2011} read: {\em ''Theories of the constitution of
bodies suppose them either to be continuous and homogeneous, or to be composed
of a finite number of distinct particles or molecules. 

In certain applications of mathematics to physical questions, it is convenient 
to suppose bodies homogeneous in order to make the quantity of matter in each 
differential element a function of the co-ordinates, but I am not aware that any
 theory of this kind has been proposed to account for the different properties
of bodies. Indeed the properties of a body supposed to be a uniform {\em plenum}
 may be affirmed dogmatically, but cannot be explained mathematically''}.

The present work is only a first step towards the mathematical theory of 'homogeneous bodies', i.e., the field theory of condensed matter. Let us hope it will motivate some critical development in condensed matter physics. 

\section*{Acknowledgment}

Let us acknowledge Grigori Vilkovisky's insight into the the value of the heat kernel for theoretical physics. We are grateful to Hermann Nicolai for encouraging new research directions at AEI. This work was partially done during the 2014 visit to Institut des Hautes \'Etudes Scientifiques. We thank Veselin Jungic at SFU's IRMACS Centre for the electronic information support. We acknowledge emergency help of Wikipedia as a guide in the information Universe.

\section*{Appendix A: Review of the Debye theory} \label{Debye}

The generating functional of the Debye theory is not dimensionless,
but it is energy (J) \cite{Debye-AdP1912}.  The Debye theory  contains one
calibration parameter, the Debye temperature, $T_{\mathrm{Debye}}$, which gives the
dimensionless variable of the theory,
\begin{equation}
\tau \equiv T_{\mathrm{Debye}}/T .
\end{equation}
The scaling function of the Debye theory has the form,
\begin{equation}
D(\tau)= \frac{3}{\tau^3} \, \int_{0}^{\tau} \mathrm{d} x \frac{x^4 \mathrm{e}^x}{(\mathrm{e}^x-1)^2}, \label{DebyeF}
\end{equation}
which gives the specific heat as,
\begin{equation}
C_{\mathrm{Debye}} = 3 k_{\mathsf{B}} N_{\mathsf{A}} D(\tau).
\end{equation}
The overall factor of the Debye specific heat is fixed by the belief in the
exact Dulong-Petit law,
\begin{equation}
C_{\mathrm{Debye}} = 3 k_{\mathsf{B}} N_{\mathsf{A}},\  \tau \rightarrow 0 .
\end{equation}
The low temperature asymptotics
\begin{equation}
C_{\mathrm{Debye}} = (12 \pi^4/5)  k_{\mathsf{B}} N_{\mathsf{A}} / \tau^3,\ 
\tau \rightarrow \infty .
\label{DebyeLT}
\end{equation}
and $ C_{\mathrm{Debye}} $ is zero at $T=0$.

The Debye temperature is an experimentally determined constant for every material.
It is found from the low-T limit of the specific heat (\ref{DebyeLT}) as, 
\begin{equation}
T_{\mathrm{Debye}} = (12 \pi^4/5)^{1/3} T (R/C_M)^{1/3}.  \label{DebyeT}
\end{equation}
 The Debye temperature is related to the elasticity
constants through the sound
velocities \cite{Debye-AdP1912,Gopal-book1966,Kittel-book1996},
\begin{equation}
T_{\mathrm{Debye}} = (\hbar v_m / k_{\mathsf{B}}) 
(6 \pi^2 N/ \mathcal{V})^{1/3}. \label{DebyeTel}
\end{equation}
The general deficiency of the Debye theory is its single characteristic
temperature that could not be physically realistic for describing specific
heats, as is recognized in the literature. With one Debye temperature, there can be
only one velocity. It is the average velocity $v_m$ of the longitudinal and
transverse velocities, 
\begin{equation}
v_m = \Big( 1/3 (1/{v_l^3} + 2/{v_t^3})\Big)^{-1/3}.
\end{equation}
In fact, at low temperatures, this expression is equivalent to using the
transverse velocity: $v_t$ is typically 1.5-2.0 times smaller than $v_l$, so,
$v_m \propto v_t$. This fact may explain why the values of the implied
Debye temperature (\ref{DebyeT}) and the elastic Debye temperature
(\ref{DebyeTel}) are so close.

It is described in the literature \cite{Gopal-book1966,Kittel-book1996}, that the
Debye theory of specific heat works relatively well at temperatures below $T <
T_{\mathrm{Debye}}/50$ and above $T>T_{\mathrm{Debye}}/2$, with different values for the Debye temperature in each range. Thus, it does not work for the large and important range of temperatures. Above we have argued that the Debye theory also fails to work at the high and low temperature asymptotics. 

The reason for poor performance of the Debye theory is its two intrinsic
problems. First, in order to deal with the divergence of total energy in the
high frequency limit, the continuous medium model based on elasticity theory was
replaced by the discrete model of quantum oscillators, which were
identified with the lattice nodes \cite{Kittel-book1996,Debye-AdP1912}. Here
Debye used Einstein's idea \cite{Einstein-AdP1907} to explain the Dulong-Petit law by postulating that the number of quantum oscillators is equal to the number of lattice nodes of a crystal lattice, $N$. Then he equated the total energy of these oscillators,
found according to the equipartition theorem, to the total energy of elastic
standing waves, in the volume ${\mathcal V}$, up to certain maximum frequency,
$\nu_{\mathrm{max}}$. He derived from the elasticity theory that the {\em total}
energy of elastic waves is proportional to the third power of this maximum
frequency \cite{Debye-AdP1912},
\begin{equation}
3 N  = \nu_{\mathrm{max}}^3 {\mathcal V} {\mathsf{F}},  \label{maxf}
\end{equation} 
where ${\mathsf{F}}$ is given by the elasticity parameters. This equation
enforces the fixed upper limit of a body's elastic spectrum. The maximum value
$\nu_{\mathrm{max}}$ obtained from (\ref{maxf}) serves as the upper cut-off of
the frequency integral. 

However, this explanation could not have been acceptable, because the Debye theory is a field model, while the discrete model of a crystal lattice presents an alternative
description of the same physical phenomena. This creates a mix-up of two entirely different formalisms, and the resulting hybrid theory is not self-consistent. Second, in his derivation of specific heat P. Debye computed an integral of the density of eigen-states over frequency. Thereby, he replaced the discrete modes by a continuous variable. However, in the low temperature regime (\ref{DebyeLT}), the deviation from the true discrete distribution inside a compact domain can be significant. In fact, this substitution procedure in general is ambiguous and serves as a cause of confusion. 

\section*{Appendix B. Specific heat of solid argon}  \label{argon}

The popular solid state physics textbook of Charles Kittel
\cite{Kittel-book1996} gives a graphic proof of the $T^3$ law by plotting the
specific heat of solid argon vs. $T^3$, Fig. 9, p. 125.  The figure is given with no
reference, but the names of experimental data's authors let us trace its published
source, which is the work of Leonard Finegold and Norman Phillips
\cite{Finegold-PR1969}. These experiments were done, at temperatures between 0.4
and 12K,  with the heat pulse technique, i.e., by determining the speed of heat
propagation in matter, as different from the calorimeter measurements. Beside
the crystal lattice of solid argon is face-centered cubic, not the diamond one
as in the examples above. However, due to significance given to this
plot in the literature we still analyzed these references and the specific heat
data for solid argon.

Using the data of Ref.~\cite{Finegold-PR1969}, the graph of $C_M$ vs. $T^3$, similar to Fig. 9 of Chap. 5 in the textbook \cite{Kittel-book1996}, can be plotted, and it does look like, at the first glance, as a straight line. However, this figure is really deceiving, because the use of such a scale on the horizontal axis does not reveal the data's power law.
This apparently straight line is determined by a few points at higher
temperatures $7\,K < T <11\,K$, while nothing quantitative can be concluded
about the majority of data points in the few Kelvin region of interest. 

\begin{figure}[h!]
\caption{Low temperature range for $C_M$ of solid argon, \cite{Finegold-PR1969}}
  \centering
\includegraphics[scale=0.5]{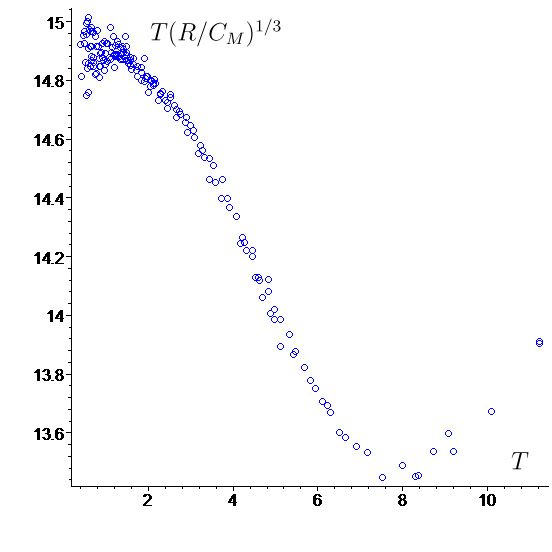}
\label{argondata}
\end{figure}

We performed on these data the same analysis as done for the carbon group
elements, and the conclusions are the same.  If the statistical analysis is done
with the raw data, $C_M(T)$, it confirms the quartic law at lower
temperatures, but the cubic law is not statistically excluded. Using the plot,
$T(R/C_M)^{1/3}$ vs. $T$, Fig.~\ref{argondata}, the characteristic temperature of solid argon is
found, $T_0=8.0K$. 

\begin{figure}[h!]
  \caption{Power law for the specific heat of solid argon, \cite{Finegold-PR1969}}
  \centering
\includegraphics[scale=0.5]{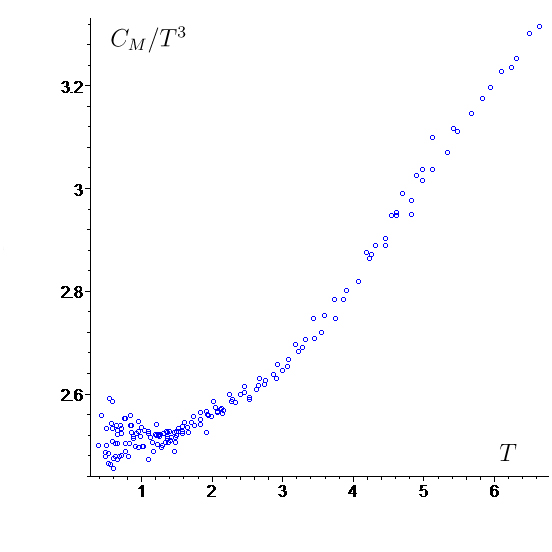}
\label{Cvargon}
\end{figure}

If we plot  $C_M/T^3$ vs. $T$, Fig.~\ref{Cvargon}, as done by L. Finegold and N. Phillips \cite{Finegold-PR1969}, we do not get a horizontal line predicted by the $T^3$-law, neither Finegold and Phillips did. 
This graph, Fig.~\ref{Cvargon} (Fig.~2 in Ref.~\cite{Finegold-PR1969}), varies significantly, likewise the krypton data's plot \cite{Finegold-PR1969}. The data points below 1.5 K should be excluded due to apparent experimental uncertainties. In the remaining range of temperatures, from 1.5 to 8 K,  this graph Fig.~\ref{Cvargon} is almost a linear function, $C_M/T^3=a\, T$, which implies {\em the $T^4$ power law} in the specific heat data of solid argon (as well of solid krypton).

These experimental data are similar to and improve on the results of J.A.
Morrison's group \cite{Flubacher-PPS1961}, as compared in
\cite{Finegold-PR1969}. The experimentalists, being unable to explain
these discrepancies with the theory, concluded their 1969 paper by calling
for better theories \cite{Finegold-PR1969}: {\em ''whereas some years ago the theories
were more advanced than the experimental data available, now the need is for a
better understanding of the interatomic forces as well as for improved
anharmonic theories''}. The old experimental data are still quite
sufficient for modern use, but instead of building a tower of improvements upon
the old theory we suggest to explore different theoretical ideas.

\pagebreak
\section*{Appendix C. Specific heat of germanium, \cite{Flubacher-PM1959}}
\label{appendixB}
The temperature is given in the units of Kelvin, and the specific heat
of germanium is converted from calories of the original table to Joules of the
SI system, $J\,mol^{-1} \, K^{-1}$.

\begin{figure}[h!]
  \caption{Specific heat of germanium, \cite{Flubacher-PM1959}}
  \centering
\includegraphics[scale=0.5]{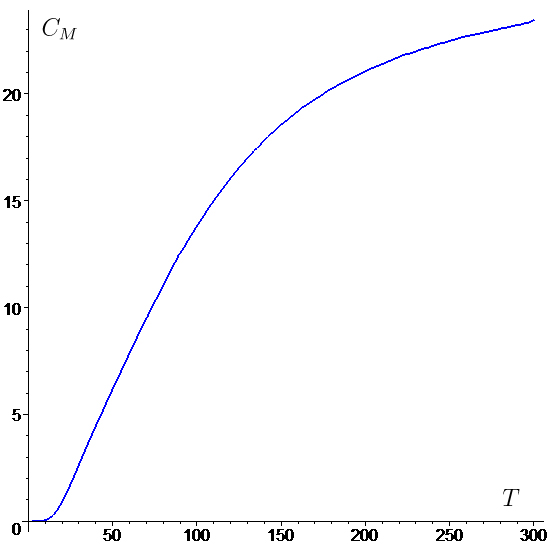}
\label{CvGe}
\end{figure}
\pagebreak
\thispagestyle{empty}
\begin{table}[!ht]
\caption{Specific heat of germanium, \cite{Flubacher-PM1959}}
  \centering
\begin{tabular}{llllllllll}
\hline
$T$ & $C_M$ & $T$ & $C_M$ & $T$ & $C_M$ &  $T$& $C_M$&  $T$& $C_M$\\
\hline
2.461	&	0.000537	&	13.008	&	0.1784	&	40.46	&	4.548	&	95.46	&	13.242	&	200.02	&	 21.062 \\
2.675	&	0.0007389	&	13.478	&	0.2069	&	42.43	&	4.8869	&	97.17	&	13.456	&	203.54	&	 21.192 \\
2.971	&	0.0009778	&	14.002	&	0.2439	&	44.29	&	5.1882	&	98.76	&	13.665	&	207.02	&	 21.301 \\
3.175	&	0.001259	&	14.536	&	0.2831	&	45.89	&	5.4894	&	100.27	&	13.845	&	210.49	&	 21.418 \\
3.481	&	0.001639	&	15.002	&	0.3200	&	47.87	&	5.8283	&	103.23	&	14.192	&	213.92	&	 21.527 \\
3.713	&	0.002040	&	15.192	&	0.3394	&	49.37	&	6.0793	&	106.87	&	14.652	&	217.34	&	 21.644 \\
3.961	&	0.002411	&	15.579	&	0.3693	&	51.90	&	6.5145	&	110.39	&	15.046	&	222.82	&	 21.836 \\
4.364	&	0.00332	&	15.945	&	0.4081	&	53.59	&	6.7864	&	113.82	&	15.406	&	226.53	&	 21.891 \\
4.471	&	0.003499	&	15.961	&	0.4112	&	55.63	&	7.1254	&	117.17	&	15.765	&	230.21	&	 21.991 \\
4.814	&	0.004506	&	16.578	&	0.4715	&	57.15	&	7.3848	&	120.44	&	16.104	&	233.87	&	 22.117 \\
4.963	&	0.004895	&	16.979	&	0.5151	&	59.02	&	7.6986	&	123.84	&	16.435	&	237.50	&	 22.171 \\
5.283	&	0.006050	&	17.579	&	0.5866	&	60.39	&	7.9287	&	127.36	&	16.761	&	241.11	&	 22.280 \\
5.503	&	0.006845	&	17.588	&	0.5853	&	62.35	&	8.2592	&	130.81	&	17.067	&	244.70	&	 22.380 \\
5.774	&	0.008012	&	18.051	&	0.6418	&	63.76	&	8.4893	&	134.18	&	17.364	&	246.4	&	 22.380 \\
6.024	&	0.009155	&	18.493	&	0.7008	&	65.81	&	8.8324	&	135.93	&	17.481	&	249.98	&	 22.485 \\
6.284	&	0.01051	&	18.628	&	0.7117	&	67.30	&	9.0751	&	139.22	&	17.761	&	253.51	&	 22.552 \\
6.506	&	0.01172	&	19.041	&	0.7640	&	69.22	&	9.3680	&	142.42	&	18.025	&	257.02	&	 22.656 \\
6.784	&	0.01359	&	19.632	&	0.8481	&	70.73	&	9.5939	&	145.75	&	18.255	&	262.81	&	 22.748 \\
7.131	&	0.01609	&	20.233	&	0.9439	&	72.54	&	9.8868	&	149.23	&	18.523	&	268.85	&	 22.849 \\ 
7.51	&	0.01953	&	21.449	&	1.1334	&	74.07	&	10.130	&	152.66	&	18.723	&	272.84	&	 22.920 \\
7.974	&	0.02409	&	23.186	&	1.4192	&	76.00	&	10.422	&	156.05	&	18.949	&	276.81	&	 22.987 \\
8.465	&	0.03021	&	24.548	&	1.6527	&	77.45	&	10.653	&	159.39	&	19.150	&	280.72	&	 23.054 \\
8.948	&	0.03745	&	25.86	&	1.8845	&	79.38	&	10.950	&	162.69	&	19.372	&	280.90	&	 23.075 \\
9.434	&	0.04619	&	27.22	&	2.1313	&	81.09	&	11.217	&	166.06	&	19.531	&	284.80	&	 23.117 \\
10.006	&	0.05820	&	28.46	&	2.356	&	82.75	&	11.477	&	169.50	&	19.715	&	286.55	&	 23.167 \\
10.442	&	0.07025	&	29.87	&	2.6221	&	84.36	&	11.719	&	172.90	&	19.887	&	288.66	&	 23.205 \\
11.015	&	0.08799	&	31.16	&	2.8614	&	85.97	&	11.962	&	176.26	&	20.067	&	292.38	&	 23.251 \\
11.458	&	0.1037	&	32.68	&	3.1418	&	87.50	&	12.180	&	178.49	&	20.184	&	292.48	&	 23.263 \\
12.011	&	0.1272	&	33.91	&	3.3677	&	89.06	&	12.439	&	183.82	&	20.418	&	296.23	&	 23.351 \\
12.366	&	0.1469	&	35.72	&	3.7020	&	90.70	&	12.615	&	189.31	&	20.631	&	296.24	&	 23.313 \\
12.471	&	0.1489	&	36.93	&	3.9229	&	92.22	&	12.791	&	192.91	&	20.782	&	300.01	&	  23.468 \\
12.902	&	0.1752	&	39.00	&	4.2928	&	93.98	&	13.054	&	196.48	&	20.916	&	300.05	&	 23.430 \\
\label{Gedata}
\end{tabular}
\end{table}

\end{document}